\documentclass{article}
\usepackage{epsfig}
\usepackage{amssymb}
\usepackage{amsmath}
\usepackage{amsfonts}
\newcommand \be{\begin{eqnarray}}
\newcommand \ee{\end{eqnarray}}
\begin{document}
\begin{center}
{\bf Efimov-physics and Beyond}\\
\bigskip
\bigskip
H. S. K\"ohler \footnote{e-mail: kohlers@u.arizona.edu} \\
{\em Physics Department, University of Arizona, Tucson, Arizona
85721,USA}\\
\end{center}
\date{\today}

\begin{abstract}
Efimov physics relates to 3-body systems with large 
2-body scattering lengths $a_s$ and small effective ranges $r_s$. 
For many systems in
nature  the assumption of a small effective range is not valid.
The present report shows 
binding energies $E_3$ of three identical bosons calculated with 2-body
potentials that are fitted to scattering data  and  momentum cut-offs $\Lambda$
by inverse scattering. Results agree with previous works in the case of
$r_s \ll a_s$. 

While energies diverge with $\Lambda$ for $r_s=0$, they converge for
$r_s>0$ when $\Lambda>\sim 10/r_s$. With $a_s^{-1}=0$ the converged
energies are given by $E^{(n)}_3 =C^{(n)}_0 r_s^{-2}$  with $n$
labelling the energy-branch and calculated values
$C^{(0)}_0=0.77, C^{(1)}_0=.0028.$ This gives a ratio  $\sim 278$ thus
differing from  the value $\sim 515$ in the Efimov case. 

Efimov's angular dependent function is calculated.
Good agreement with previous works is  obtained for $r_s \ll a_s$. 
With the increased values of $r_s$ the shallow states  still appear Efimov-like. 
For deeper states the angular dependence differs but is  independent 
of $r_s$.
\end{abstract}

\section{Introduction}
Systems involving particles with 2-body interactions at or close to the
Unitary limit have become of specific interest in the physics community during the
last several years for reasons that have repeatedly been pointed out in
numerous publications related to both atomic and sub-atomic problems. 
The Unitary limit is here defined as being that for which the scattering length
$a_s$ and the effective range $r_s$ is infinite and zero respectively.

In this regard the bound system of three bosons is of special interest.
This was 
first brought to the attention by the works of Thomas \cite{tho35} and of
Vitali Efimov\cite{eff70,ama72}.  
 Bosons interacting with a resonance in the
2-body state (i.e.  $a_s^{-1}\sim 0$)  result in a strongly bound 
three-body system \it and \rm with a spectrum  of loosely bound excited states.

The literature has brought to the attention several systems in nature for
which these findings are relevant. 

On the theoretical side the application of Effective Field Theory (EFT)
methods have been proven very 
powerful\cite{bed99,bra03,bra06,bra07,moh08,ham11} for a theoretical
interpretation.
The main focus has been on short-ranged potentials at or near the Unitary
limit for which $r_s\ll a_s$, but more recent EFT  publications include
effective range corrections.\cite{ji11,pla06}

A rather different approach is used here. The inverse scattering formalism for
separable potentials is used to construct
potentials as functions of the scattering parameters as well as
2-body binding energy, and renormalised by a cut-off in momentum-space.
The potential in the
Unitary limit, a renormalised $\delta$-function (in coordinate)
space, is of specific interest. 
The results presented below  agree with previous Efimov
physics calculations for $r_s=0.$. The $r_s>0$ results do in general show
some qualitative differences.

The present investigation focuses on  Efimov 
energies and the Efimov function,  dependence on 
scattering length and in particular on the effective range.
Some preliminary results were given in ref.\cite{hsk10b}

Scaling properties are a main thread in this work.

In  Section 2 is found a presentation of the necessary tools which are the
Faddeev equation and the inverse scattering method.
Section 3 show results of numerical calculations in 4 subsections with 8 Figures.
Section 4 is a summary and some discussion of the results. 
Some relations regarding off-shell scattering and three-body forces
are shown in  Appendix A, while Appendix B shows an important  relation 
pertaining to scattering and the separable interaction in the Unitary
limit.

\section{Formalism}
\subsection{Two-body Separable Interaction}
The input for the calculations in this report are scattering phase-shifts
$\delta(k)$
related to the scattering length $a_s$ and an effective range $r_s$
by
 \begin{equation} 
k\cot\delta(k)=-\frac{1}{a_s}+\frac{1}{2}r_sk^2
\label{phase}
\end{equation}
A momentum cut-off $\Lambda$ will also be defined so that
$$\delta(k>\Lambda)=0$$.

With the scattering parameters chosen so that the phase-shifts do not
change sign, as will be the case here, they can be reproduced by a rank-1
separable potential

\begin{equation}
V(k,k')=-v(k)v(k')
\label{V}
\end{equation}
that is obtained by inverse scattering from\cite{tab69}
\begin{equation}
v^{2}(k)= \frac{4\pi}{k}\sin \delta (k)|D(k^{2})|
\label{v2}
\end{equation}
where
\begin{equation}
D(k^{2})=\frac{k^2+E_B}{k^2}\exp\left[\frac{2}{\pi}{\cal
P}\int_{0}^{\Lambda}
\frac{k'\delta(k')}{k^{2}-k'^{2}}dk' \right]
\label{D}
\end{equation}
where ${\cal P}$ denotes the principal value.
The interaction is fully defined by the phase-shifts and
the two-body binding energy $E_B$. Note that the two fits, to scattering
and to binding energy are independent.   The equation 
\begin{equation}
\sqrt{E_B}=(1-\sqrt{1-2r_sa_s^{-1}})\frac{1}{r_s}
\label{EB}
\end{equation}
relates $E_B$ to [$a_s,r_s$]  but only for $r_s\ll a_s$ and it is not useful
for the calculations below.
In the limit $r_s=0$ it reduces to
$$E_B=a_s^{-2}$$
which will be used also when $r_s\neq 0$ together with $E_B=0$ for $a_s<0$. 
This simplifies the comparison with the $r_s=0$ calculations in
these preliminary calculations. The major importance is that 
the 2-body system \it has \rm a bound
state. 
\footnote{Calculations showed no qualitative
difference if using eq. (\ref{EB}).}

When dealing with some specific physical system one (in general) 
knows or assumes some binding-energy, but this is not the
case here. The expression (\ref{EB}) for the binding energy does not
introduce another parameter in the theory, being a function of $a_s$ and
$r_s$ only, but if in some specific case one would tune the binding enrgy
say by some shape-parameter there would be another parameter to consider.

\subsubsection{Unitary Limit}
A well-known reason for \it assuming \rm  a separable interaction in 3- and other 
many-body works is
that it in general simplifies the formalism as well as the 
numerical calculations. There is however a particular case where  the
interaction IS separable. Namely that when the 2-body system has a bound
state at or close to zero\cite{gebrown}, which is the case in the Unitary
limit.

The principal value integration in eq. (\ref{D}) can be done analytically
for $\delta(k)$=constant.  The Unitary limit is a special case with  
$\delta(k)=\frac{\pi}{2}$, for which 
the rank-1 separable potential  $v_u(k)$ is given by \cite{hsk10}

\begin{equation}
v_u^{2}(k)= - \frac{4\pi}{(\Lambda^{2}-k^{2})^{\frac{1}{2}}}
\label{vpi2}
\end{equation}
If $\Lambda \gg k$ one finds
\begin{equation}
v_u^{2}(k)\rightarrow - \frac{4\pi}{\Lambda}
\label{v2ka}
\end{equation}
In this limit, \it but only in this limit \rm ,
the unitary interaction is independent of momentum and a
$\delta$-function in coordinate space with the strength  inversely
proportional to the cut-off. But for the finite values of $\Lambda$  needed for
computations, renormalisation is required resulting in $v_u$ as given by eq.
(\ref{vpi2}).
Note that there is then an abrupt increase in strength and a singularity 
as $k\rightarrow \Lambda$. This is required to 
preserve the condition $\delta=\frac{\pi}{2}$ for all $k\leq \Lambda$,
as required for the Unitary limit. (See Appendix B.)

\begin{figure}
\centerline{
\psfig{figure=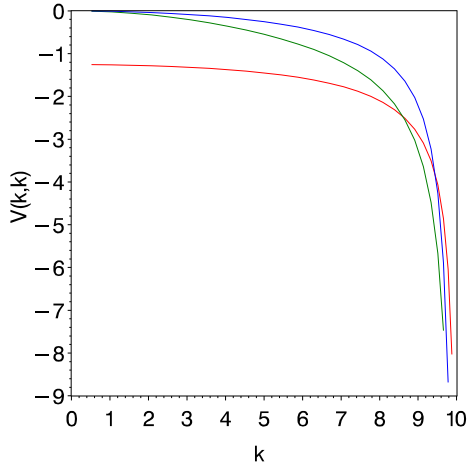,width=7cm,angle=0}
}
\vspace{.0in}
\caption{
Lowest curve (red on-line) shows the diagonal of the Unitary interaction in
momentum-space.
The two other curves are, from below, the potentials for $a_s=10$ (green
on-line) and for $a_s=20$ (blue on-line). All three are with the effective
range $r_s=0$. 
}
\label{efvsep}
\end{figure}

The singularity can cause numerical problems that however can be largely
overcome by proper computational  methods.
The substitution $$k=\Lambda \sin\theta,$$ with $\theta$ the new variable,
can for example be helpful yielding
\begin{equation}
v_u^{2}(\theta)= - \frac{4\pi}{\Lambda \cos\theta}
\label{vtheta}
\end{equation}
which is used successfully in the computations. \footnote{This is in particular
useful because $dk \rightarrow \Lambda \cos\theta d\theta$ which can 
eliminate the $\cos\theta$ in the denominator of (\ref{vtheta})}.
Fig. \ref{efvsep} shows the Unitary interaction (lowest curve) and for
comparison two other potentials with $a_s=10$ and $20$ respectively. 

\subsection{The Many-body System}
A rank-1 potential is often sufficient to fit the on-shell data at low energy 
as is done in this report. The off-shell and off-diagonal parts of the
T-matrix is then  fixed by default. 

On-shell  relates to the asymptotic form of the two-body scattering 
wave-function, while  off-shell relates to the interior wave-function,
i.e. to correlations in the two-body system.  One may however like to change 
the off-shell behaviour.
This can be done by extending to a rank-2 (or higher)
potential (which implies introducing another parameter), while preserving the
on-shell fit.  This provides a practical tool for exploring off-shell 
effects on many-body properties. 
\footnote{It  was for example used by the author in earlier work on the triton.
\cite{hsk09}}

The 3-boson system collapses in the Unitary limit
(the energy diverges) with  $\Lambda$. This is well-known. 
In ref.\cite{bed99} this collapse was remedied by adding a 3-body force. 
It was shown in ref.\cite{hsk10b} that the same
could be accomplished by a change in off-shell scattering with a rank-2
potential as described above.

In each of these methods  the  high energy 
(ultraviolet) region of the interaction was modified;
in one case by a three-body force in the other by
a change in off-shell scattering. 

Although the two methods have a similar 'end-result' it is important 
to understand that in each case there is a different physics involved. 
(Appendix A.)

Findings below  show that the effective range also has
the effect of preventing a collapse. This is not completely un-expected as it
explicitly introduces a finite range parameter into the system.

\subsection{Faddeev  Equation}
The Faddeev three-boson equation for a spin-independent rank-1 separable 
potential is given by 
 \begin{equation} 
 \chi(q)=\frac{2}{{\cal I}(E_3-\frac{3}{4}q^2)}\int_{0}^{2\Lambda} 
 \frac{v(|{\bf k}+\frac{1}{2}{\bf q}|)v(|{\bf q}+\frac{1}{2}{\bf k}|)}
 {q^2+{\bf q}\cdot {\bf k}+k^2-E_3}\chi(k)d{\bf k}
\label{faddeev}
\end{equation}
with 
\begin{equation}
{\cal I}(s)=1+\frac{1}{2\pi^2}\int_{0}^{\Lambda}
v^{2}(k)(s-k^2)^{-1} k^{2}dk
\label{DG}
\end{equation}
With the phase-shifts defined, as announced above, in terms of  two 
scattering parameters and the
cut-off $\Lambda$ one will have $E_3=E_3(a_s,r_s,\Lambda)$.
In the present investigation the scattering parameters $[a_s,r_s]$ are
considered \it internal \rm parameters defining the 3-body system, while
$[\Lambda]$ is considered an \it external \rm parameter if possible chosen larger
than the maximum momenta of the bound particles. This means in
general that one has to choose $\Lambda >1/R$ with
$R$ being  the physical size of the system. This size is a 'functional' of the
interaction (the internal parameters) and so is therefore $\Lambda$. As an
example, the three-body system collapses in the Unitary limit, with
$\Lambda \rightarrow \infty$ , a well-known divergence.

Three separate cases will be considered here with results presented in
three separate sections. The first will be the Unitary
limit i.e.  both $r_s$ 
and $a_s^{-1}=0$. The second is $a_s^{-1} \neq  0$ while $r_s=0$ and the third the
more general case, both $r_s$ and $a_s^{-1} \neq 0$. The purpose of the present
investigation is in particular to investigate the latter case, with $r_s > 0$.

\section{Numerical Results}
\subsection{Unitary Limit}

The Unitary limit is a special case.
After a change of variables (see above),  $k \rightarrow \Lambda \sin\theta,$ 
the function  ${\cal I}(s)$, eq.  (\ref{DG}), becomes   
\begin{equation}
{\cal I}(s)=1+\frac{2}{\pi}{\cal P}\int_{0}^{\frac{\pi}{2}}
\sin^2\theta(s-\sin^2\theta)^{-1} d\theta
\label{UDG}
\end{equation}
The integral, done analytically for $s<0$ yields
\begin{equation}
{\cal I}(s)=-\frac{s}{\sqrt{s^2-s}}
\label{UDG1}
\end{equation}
(See Appendix B for $s>0$).
The only free parameter in the Unitary limiit is $\Lambda$ so that one has
$E_3=E_3(\Lambda)$.
It is then convenient in this case to choose
momenta and energies in units of $\Lambda$ and $\Lambda^2$ respectively by
substitutions $k\rightarrow k\Lambda$ $q\rightarrow q\Lambda$ and 
$E_3=E_u\Lambda^2$ in eq. (\ref{faddeev}) to get with eq. (\ref{UDG1}) and
$s=E_u-\frac{3}{4}q^2$

\begin{equation}
\chi(q)=-\frac{2\sqrt{s^2-s}}{s}
\int_{0}^{2} 
\frac{v_u(|{\bf k}+\frac{1}{2}{\bf q}|)v_u(|{\bf q}+\frac{1}{2}{\bf k}|)}
{q^2+{\bf q}\cdot {\bf k}+k^2-E_u}\chi(k)d{\bf k}
\label{unifadd}
\end{equation}
With $n$ labelling a specific state with energy $E_u^{(n)}$
the eigenvalue spectrum in this Unitary limit was calculated 
with  the result (in units of $\Lambda^2$):
$$E_u^{(0)}=-.1325, \;\;\;E_u^{(1)}=-.0002520, \;\:\: E_u^{(2)}=-.0000004568$$
One finds
 $$E_u^{(0)}/E_u^{(1)}=515$$ agreeing with the Efimov result while
 $$E_u^{(1)}/E_u^{(2)}=630$$
 This latter discrepancy is ascribed to computational inaccuracies that
 increases for the
 smaller energy.  ($E_u^{(2)}=-.0000004893$ would be the 'correct' value.)

\subsection{Non-zero Scattering Length but Zero Effective Range}
In this second scenario with $r_s=0$ while $a_s \neq 0$ one will have 
$E_3=E_3(a_s,\Lambda)$. This case has been the focus of numerous
investigations
since the early works of Thomas and Efimov; in particular as relates to
experimental studies of cold atoms and the ability to tune the
scattering length with the aid of  Feshbach resonances.

\begin{figure}
\centerline{
\psfig{figure=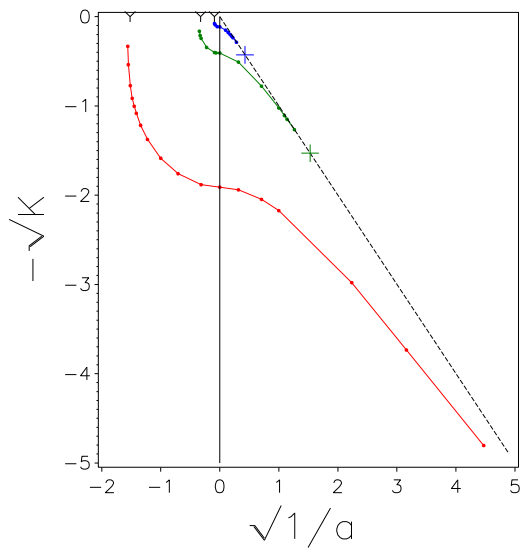,width=7cm,angle=0}
}
\vspace{.0in}
\caption{
Energy as a function of scattering length $a_s$ while the effective range is
$r_s=0$. Following convention,
$K\equiv \kappa=\sqrt{E_3}$. The lowest curve (red on-line) is for 
$n=0$. The two
other curves (green and blue respectively on-line) are for $n=1$ 
and $2$ respectively. 
The crosses on the dimer-energy line, shown by the broken line,  are 
explained in the text.  $\Lambda=10$ in this plot.
}
\label{ef12b}
\end{figure}
Results of Faddeev calculations are shown in Fig. \ref{ef12b}. 
Following convention, the figure shows $E_3^{\frac{1}{4}}$ vs 
$a_s^{-\frac{1}{2}}$ (with
appropriate signs) for three separate branches
cutting the $a_s=0$ axis at the values shown above.
With $\kappa_{*}=\sqrt{E_3(a_s^{-1}=0)}$ for each branch
it has been found, e.g. \cite{bra06}, that the energys should approach zero
for $a'_s \sim -1.56 \kappa_{*}^{-1}$.  These scattering lengths are
indicated by a "Y" for each of the states and largely agree with the calculations.
Furthermore, the trimer and dimer energies are estimated to coincide at
$a_{*}=0.0707645086901 \kappa_{*}^{-1}$. These points are shown by crosses
in Fig. \ref{ef12b} for the two shallow states. For the lowest state the
corresponding point is predicted to be at  $a_s^{-1} \approx 51$. 

\begin{figure}
\centerline{
\psfig{figure=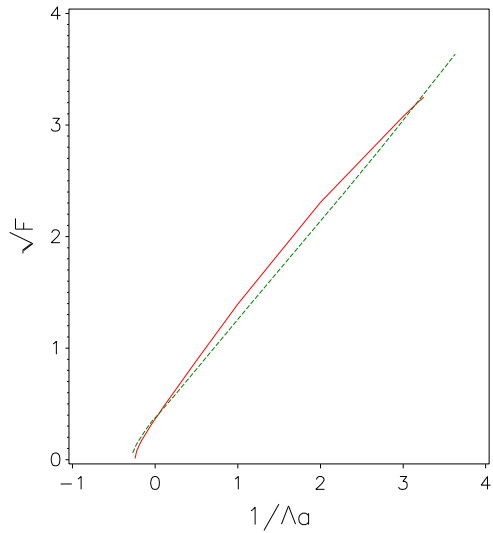,width=7cm,angle=0}
}
\vspace{.0in}
\caption{The function $\sqrt{F^{(n)}}$ is shown for the two deepest states.
The full curve  (red on-line) is for $n=0$ while the broken (green on-line)
is for $n=1$. 
This curve has been scaled up  by the Efimov factor
$22.7$  as a result of which
the two curves are close. Note also the approximately linear function of
 $1/a_s\Lambda$.
}
\label{ef6c}
\end{figure}

It is convenient to choose $\Lambda$ as the unit of momentum and  
introducing the dimensionless variable $a_s\Lambda$. The phase-shifts and
therefore the potential are then functions of this variable. So in addition
to the $\Lambda^2$ factor the energy will also be a function of the same
variable. One will have
\begin{equation}
E^{(n)}_3=\Lambda^2F^{(n)}(1/\Lambda a_s)
\label{F}
\end{equation}
where as before $n$ labels a specific  
(Efimov) branch.  (The reason for choosing the inverse of $\Lambda a_s$ 
becomes obvious below.)
While Fig. \ref{ef12b} shows the energy vs $a_s$ for $\Lambda=10$
the energy for
any other value of $\Lambda$ and $a_s$ can be obtained from  
eq. (\ref{F}). Of interest  is then the function $F^{(n)}$.
The full line in Fig. \ref{ef6c} shows  $\sqrt{F^{(n)}(1/\Lambda a_s)}$ for $n=0$, 
the deeper state in Fig. \ref{ef12b}. It is seen that this function is to
some approximation a linear function with the slope (derivative) being
approximately one. The function $F^{(1)}$ is a (nearly) scaled copy of
$F^{(0)}$ the scaling factor being $22.7$, as shown by the
dotted line in Fig. \ref{ef6c}. The reason for displaying these functions
is for  comparison  with the similar situation when $r_s \neq 0$ 
that is shown in Sect.  3.4. 

From the functions $F^{(n)}$ one can calculate the
energies $E_3$ for any other variables $a_s$ and $\Lambda$ for each branch $n$.
These functions can of course also be calculated directly
from the Efimov Universal function, $\Delta(\xi)$. It  is related to 
the energy $E_3$ by

\begin{equation}
E^{(n)}_3=-a_s^-2+\kappa^2_{*}(e^{-\frac{2\pi}{s_0}})^{n-n_0}
e^{\frac{\Delta(\xi)}{s_0}}
\label{Ea}
\end{equation}
where $s_0\approx 1.00624$ and $n$ labels a specific Efimov branch and $n_0$
that branch for which $\kappa^2_{*}=E_3(a_s^{-1}=0)$ .
The angle $\xi$ is defined by
\begin{equation}
\tan\xi=-a_s\sqrt E^{(n)}_3
\label{xi}
\end{equation}
It is esily seen that because of the the scaling relation shown above 
$\Delta(\xi)$ is indeed a function of $\xi$ only, independent of $\Lambda$. 

$\Delta(\xi)$ was calculated with twelve
significant figures by Mohr\cite{moh08}. Earlier results by Braaten 
et al \cite{bra03} were shown parametrized in their publication.

\begin{figure}
\centerline{
\psfig{figure=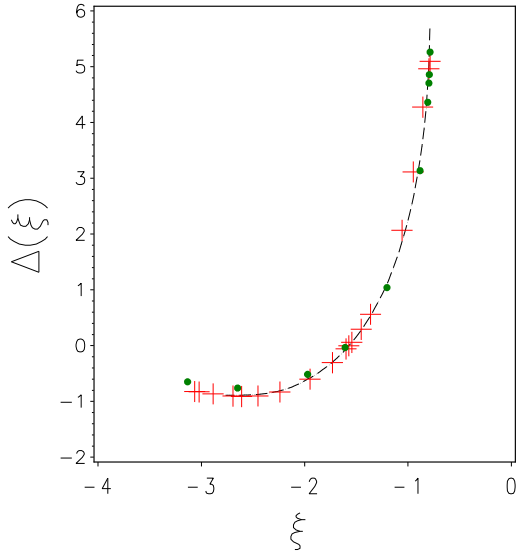,width=7cm,angle=0}
}
\vspace{.0in}
\caption{
The function $\Delta(\xi)$ is shown.
The crosses (red online) are for $n=0$  the  dots (green on-line) are 
for $n=1$.
The broken line (black on-line) shows the  results of  ref.\cite{bra03}.
}
\label{ef2b}
\end{figure}

The dots and crosses in
Fig. \ref{ef2b} show $\Delta(\xi)$ calculated from eq.(\ref{Ea}) using 
the same energys as used for  Fig. \ref{ef12b}. 
The crosses  refer to the
energy $E^{(0)}$, the deepest state,  and the dots  to $E^{(1)}$, the next
shallower state.
The broken line is from the parametrisation given in ref \cite{bra03}. 
The present calculations are not of high-precision but the difference 
between our result and those of ref. \cite{bra03} are still less
than a few percent.

\subsection{Non-Zero Effective Range and Zero Scattering Length}
After having verified that the $r_s=0$ calculations with the
separable interactions agree with previous works the rsesults with
$r_s\neq 0$ can now be presented.

The scenario changes drastically with the introduction of a non-zero
effective range. As a preliminary to the more general case the scattering
length is first assumed to be infinite. The remaining variables are then $r_s$ and
$\Lambda$. While in the previously considered case $E_3 \propto \Lambda^2$
associated with a collapse  of the system (in coordinate space),
the range-parameter  $r_s$ now provides some finite size $R$ of the system under
consideration. As is to be expected
computations of $E_3$ as a function of $\Lambda$ now show a convergence
for  $\Lambda>1/R$. With $\Lambda$ providing a momentum scale 
and following arguments above one can  expect a relation of the following form
\begin{equation}
E^{(n)}_3=\Lambda^2F^{(n)}_0(r_s \Lambda)
\label{Ers}
\end{equation}
In order for  convergence  one must then have
\begin{equation}
F^{(n)}_0(r_s \Lambda) \rightarrow \frac{1}{(r_s\Lambda)^2} 
\end{equation}
for large $\Lambda$.
The energy of the three body system will then scale with $r_s$ as
$$E^{(n)}_3=\frac{C^{(n)}}{r_s^2}$$
for $\Lambda$ large.
The coefficients $C^{(n)}$ were calculated to get  $C^{(0)}=0.770$ and 
$C^{(1)}=0.00298$.  The ratio $C^{(0)}/C^{(1)}\simeq 278$ differs from 
the related ratio of $515$ for Efimov states. 

\begin{figure}
\centerline{
\psfig{figure=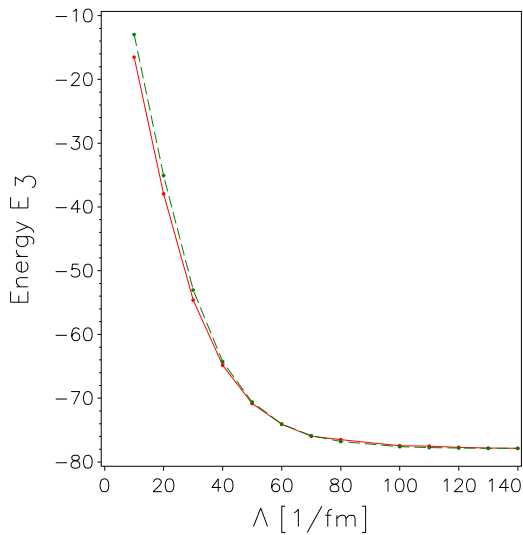,width=7cm,angle=0}
}
\vspace{.0in}
\caption{
The energies $E^{(0)}_3$ (red on-line) and $E^{(1)}_3$ (green on-line) are shown 
here as a function of the
momentum cut-off $\Lambda$ with the effective range being $r_s=0.1$ while
$a_s^{-1}=0$. 
The energy $E^{(1)}_3$ is multiplied with the
factor "278" (see text) and the two energies practically overlap.
The energy for any other [$\Lambda,r_s$] is obtained by the scaling shown in
the text. Of particular interest is the asymptotic value, i.e.  $\Lambda
>\Lambda_c$, that scales as $r_s^{-2}$. 
}
\label{ef5}
\end{figure}

Fig. \ref{ef5} shows $E^{(0)}_3$ and $278*E^{(1)}_3$ as a function of 
$\Lambda$ for $r_s=0.1$. The energy for any other values of the parameters
$r_s \neq 0$ and $\Lambda$ can be calculated by the help of  eq. (\ref{Ers}). 
Calculations have shown that the energy converges to the asymptotic 
value at a critical value $$\Lambda_c\sim 10/r_s$$ as in Fig.\ref{ef5}
where one finds $\Lambda_c\sim 100$ for $r_s=0.1$.
\footnote{Fig. 3 in ref.\cite{hsk10b} shows $E$ vs $\Lambda$ for
$r_s=.0,.03,.05$ and $.1$ with convergence at $\Lambda_c \sim 10/r_s$.
The convergence $\Lambda_c$, as a function of $r_s$ is also  seen. The curves
for $a_s^{-1}=0$ (excluding the $r_s=0$) are related by the scaling 
in eq. (\ref{Ers})}.

\subsection{General Case}
The main inspiration for the present work has been to find 
the effect of a non-zero 
effective range $r_s>0$ together with an inverse scattering 
length $a_s^{-1}\neq 0$ as opposed to the case when $r_s=0$. 

There is an additional factor to address here, the dimer (two-body) 
binding energy.  In the inverse scattering theory with separable 
interactions it is chosen separate, independent of the fit to scattering data. 
Following the discussions in Sect. 2.1 the choice here is $E_B=a_s^{-2}$. 
Any other choice might
change our results but (probably) only in a quantative way. In cases of
real physical systems one would of course choose the appropriate dimer binding 
energy assumed known for the particular system under consideration. One
consequence of this choice is that no additional parameter is introduced in
the calculations.
With $\Lambda>\Lambda_c$ (see above) the energy is then a function only of  
$a_s$ and $r_s$. Choosing momenta in units of $r_s^{-1}$ the potential will
be a function of the dimensionless parameter $a_s/r_s$.
The energy $E_3$ then has to have the form:
\begin{equation}
E^{(n)}_3=r_s^{-2}F^{(n)}_1(r_sa_s^{-1})
\label{F1}
\end{equation}
where the functions 
$F_1^{(n)}$ are  to be determined computationally.\footnote{Like
in Sect. 3.2 for the function $F$ the argument of $F_1$ is chosen as
the inverse of $a_s/r_s$.}
The scaling shown here is analogous to that shown in eq. (\ref{F}), with
$r_s$ replacing $\Lambda$  and providing the new momentum scale. 

\begin{figure}
\centerline{
\psfig{figure=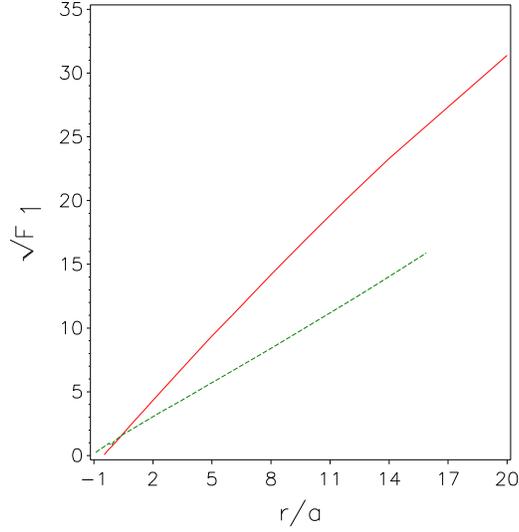,width=7cm,angle=0}
}
\vspace{.0in}
\caption{
The full curve (red on-line) shows $\sqrt{F_1^{(0)}}$ as a function of 
$r_sa_s^{-1}$. 
The broken curve (green on-line) shows $\sqrt{F_1^{(1)}}$ scaled up as in
Fig. \ref{ef6c}. 
It is seen that it (the broken)
only spans values of $r/a<1$ (note the scaling factor "22.7") so that 
Efimov physics would or might apply here.
The full line,
for $n=0$, (red on-line) includes energies for $r_s/a_s=20$  and there is
as a consequence an apparent deviation from Efimov physics here. 
This is substantiated by the results shown in Fig.  \ref{ef2c}.
}
\label{ef6ba}
\end{figure}

Fig. \ref{ef6c} showed  functions $F^{(n)}(1/\Lambda a_s)$. The similar functions 
$F^{(n)}_1(r_s/a_s)$ are shown  in Fig.  \ref{ef6ba}.
The function $F^{(1)}_1$ maps onto  $F^{(1)}$ in Fig. \ref{ef6c}, 
with a scaling factor of $\sim 3$.
So Efimov physics seems to apply for the shallow state. With regard to the
deeper state there are some qualitative similarities but there are definitely
quantitative differences.  Rather than having a slope
$\sim 1$ as in the Efimov case, Fig. \ref{ef6ba} shows $\sqrt{F^{(0)}_1}$ to have a
slope of $\sim 1.5$.

\begin{figure}
\centerline{
\psfig{figure=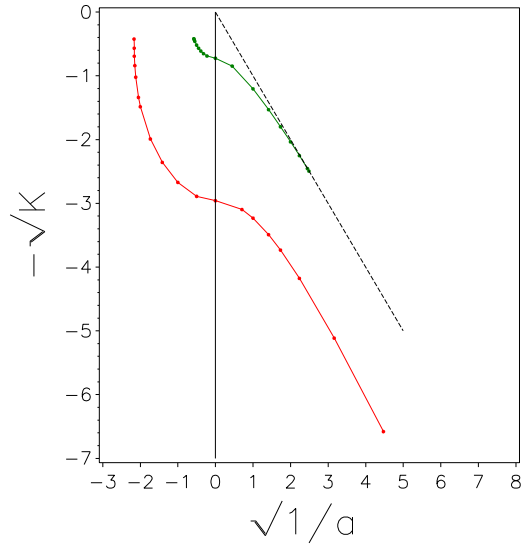,width=7cm,angle=0}
}
\vspace{.0in}
\caption{
Energy ($K$) as a function of scattering length for $r_s=0.1$. The energy
ratio between the two states at $1/a=0$ is here $\sim 278$. This factor 
is independent of $r_s$. It is approximate and improved calculations are
called for.
}
\label{ef11a}
\end{figure}

\begin{figure}
\centerline{
\psfig{figure=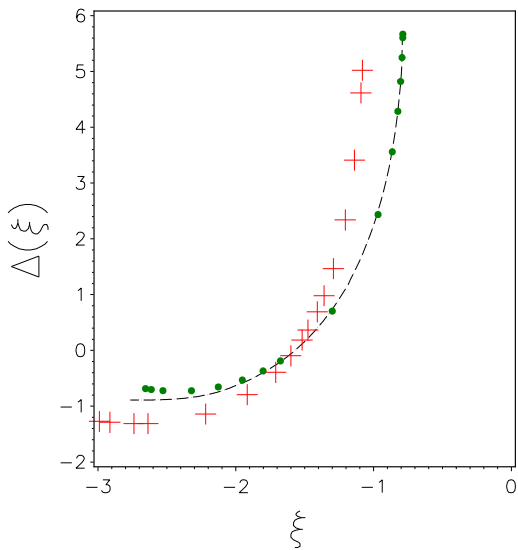,width=7cm,angle=0}
}
\vspace{.0in}
\caption{
The function $\Delta'(\xi)$ calculated from eq. (\ref{Ea}), with "515"
replaced by "278" as described in text,  for each of the
two states shown in Fig. \ref{ef11a}.
This function is independent of the value of the effective range $r_s$.
The dots (green on-line) refers to the shallow state, while the crosses
(red on-line) the deeper. The broken line (black on-line) shows the Efimov 
function using the parametrization in ref.\cite{bra03}. 
}
\label{ef2c}
\end{figure}

All the results in this section, for $r_s \neq 0$ can be obtained by
scaling from the functions $F^{(n)}_1$ shown in Fig. \ref{ef6ba}. 
As a first example is shown Fig. \ref{ef11a},  the energy
vs $a_s$ for $r_s=0.1$.  These curves are seen to be qualitatively similar to the
'Efimov-curves' shown by Fig. \ref{ef12b}. From these data a function
$\Delta'(\xi)$ was calculated using eq. (\ref{Ea}) with the result shown in 
Fig. \ref{ef2c}. 
There is a factor $(e^{-\frac{2\pi}{s_0}})^{n-n_0}=515^{n-n_0}$ in eq.
(\ref{Ea}) for $\Delta(\xi)$.
For the purpose of calculating $\Delta'(\xi)$ it was replaced by
$278^{n-n_0}$.
The function $F_1^{(0)}$ generated
the  result shown by the crosses, while $F_1^{(1)}$ the dots. The dots,
referring to the shallowest state  spans $r_s/a_s< 0.8$.
They coincide with the Efimov $\Delta(\xi)$ (the broken line), while there is a
definite difference for the deeper state derived from $F_1^{(0)}$.

The situation here is however anologous to that in sect. 3.2. There, the
consequence of the scaling expressed by eq. (\ref{F}) had as a result that
the calculation of $\Delta(\xi)$ was independent of $\Lambda$. The scaling
expressed by eq. (\ref{F1}) can formally be obtained by replacing $\Lambda$
in eq. (\ref{F}) by $r_s$. The function $\Delta'(\xi)$ is similarly
independent of $r_s$.

\section{Summary}
Bound state energies $E_3$
of the 3-boson system with 2-body interactions at or close to
the Unitary limit were calculated with the Faddeev equation in
momentum-space. Separable potentials were obtained by inverse scattering
as a function of scattering length $a_s$, effective range $r_s$ and momentum 
cut-off $\Lambda$.

In the Unitary limit ($r_s=a_s^{-1}=0$) three branches (labelled $n=1-3$)
were identified 
with energies $E_3^{(n)}$ (in units
of $\Lambda^2$),  $-.1325,-.000252,-.00000045668$ respectively.
The ratio $E_3^{(n-1)}/E_3^{(n)}\sim 515$  in agreement with theory.

With $r_s=0,a_s\neq 0$ (Sect. 3.2) the
energy for each branch scales as $E^{(n)}_3=\Lambda^2E^{(n)}(a_s\Lambda)$.
Efimov's function $\Delta(\xi)$ was calculated  separately 
for two branches ($n=1,2$) with the result shown in Fig. \ref{ef2b}
agreeing with previous works \cite{bra03,moh08}.

While the energy $\propto \Lambda^2$ for $r_s=0$ 
the situation changes when $r_s>0$.  One still finds the energy $\propto
\Lambda^2$ for small $\Lambda$ but it converges  for
$\Lambda>\Lambda_c\sim 10/r_s$ as seen in  Fig. \ref{ef5}
(Sect.3.3) for $a_s^{-1}=0$.  It was shown that the energy converges  
to $E_3{(n)}= C^{(n)} r_s^{-2}$ with $r_s^{(-1)}$ being the only remaining
momentum-scale with $\Lambda>\Lambda_c$.  The ratio $C^{(n-1)}/C^{(n)}$ 
is calculated to be $\sim 278$ differing from the corresponding  
Efimov value $\sim 515$. 

It was shown in Sect. 3.4 that 
if both $r_s>0$ and $a_s^{-1}\neq 0$ the energy can be written as
$E^{(n)}_3=r_s^{-2}F^{(n)}_1(r_sa_s^{-1})$. The function $F^{(n)}_1$
was shown and a function $\Delta'(\xi)$ was calculated analogous to 
the Efimov function $\Delta(\xi)$.
Fig. \ref{ef2c} shows it to overlap with the
Efimov for the shallow state. This might be explained by the fact that for 
this state, $r_s$ is still $< a_s^{-1}$. For the deeper state, there 
is however a definite difference even for that range of $a_s$. 
An important result following from the expression (\ref{F1}) for the energy
$E_3^{(n)}(r_s,a_s)$ is however that $\Delta'(\xi)$ is independent of
$r_s>0$ also for these states. Further investigation is asked for here.

There would (probably) be some quantative changes in the results above if
a different algorithm  were used for the binding energy as a function of
$a_s$ and $r_s$, but this all relates to a specific system, beyond the
scope of the present investigation. It would in general introduce another
parameter into the theory.

The Efimov physics, valid for $r_s/a_s\ll 1$ is  well understood and
explained by various methods. Particularly powerful is  the EFT approach
with recent and on-going research\cite{ji11,pla06} extending this method 
to finite values of the effective range.


The main purpose of this
work is to provide an alternative approach to the the problem at hand and
in particular to explore the domain of $r_s>0$. The results were obtained 
with  rank-1 separable potentials. It is to be expected that any other
potential fitted to the same scattering data would give the same, or very
similar,  results.

Although much care was taken, the computing was not of 'high precision'
and could be improved upon, but this is not expected to have an impact 
on qualitative conclusions.

\section{Appendix A}
The $\cal R$ (or T)-matrix relates the potential to observable scattering data
while the potential  in itself is not an observable.
The phase-shifts $\delta(k)$ are related to the reactance matrix $\cal R$
and indirectly to the two-body interaction $V$ by
\begin{eqnarray}
&{\cal R(\omega)}=V+VP\frac{1}{\omega-k'^2}{\cal R}K& \\
&<k|{\cal R}(\omega = k^2)|k>=\frac{1}{k}\tan\delta(k)&        \nonumber 
\label{RR}
\end{eqnarray}
As a consequence only  \it diagonal \rm elements  of the ${\cal R}$-matrix 
can be obtained from the phase-shifts and only for $\omega=k^2$. In
addition, experimental information puts an upper limit on $k$. A large part
of the matrix is  therefore  beyond direct experimental reach. Many-body 
theories do in general need more, requiring some other input, e.g. 
by potential models.

An important example is off-shell information.  Off-shell 
implies $\omega \neq k^2$. It is therefore relevant to consider
\begin{eqnarray}
\frac{\partial {\cal R}}{\partial \omega}&=&\langle
\Psi-\Phi|\Psi-\Phi\rangle =I_w   \\
\Psi&=&\Phi+\frac{1}{\omega-k'^2}V\Psi
\nonumber
\label{Psi}
\end{eqnarray}
where $\Phi$ is the unperturbed wave-function.

These relations were already used by Moszkowski and Scott.\cite{mos60,hskm07}
$I_w$ is often referred to in the literature as the wound-integral.
The 'correlated' wave-function $\Psi$ relates
to details of the interaction  (even beyond what may be available from
experiments.) 

Assume a many body system with 2-body interactions only; no 3-body (or
$n$-body, $n>2$) forces. 
Then consider two particles (i,j) interacting in this system. 
The interaction would be off-shell, $\omega \neq k^2$ , mainly because 
of binding-effects,  meaning that other 
particles (k) affect the the interaction between (i,j). So to calculate 
something, e.g the total
energy, one would have to sum over (i,j,k) that 'looks' as if one is
dealing with a 3-body \it force \rm, although 2-body only was assumed.
It is then more appropriate to say one is dealing with 3-body \it terms
\rm not \it forces \rm. 
Whether the 3-body term is 'important' or not is another
question. There are of course in principle also 4,5 etc body terms.
That depends on the system in question including density, temperature etc.

The above is to illustrate the popular statement of equivalence
between  the off-shell effect and  that of a 3-body force, even though
they  have a different origin. The first is a medium-effect, the second
relates to internal degress of freedom of the particles. Either or both
may be relevant for a specific system in nature. 
In either case there will for example be the same energy vs density
functional.

\section{Appendix B}
The scattering phaseshift is related to the reactance matrix in Appendix A.
For a rank-1 separable potential one finds
\begin{equation}
<k|{\cal R}(\omega = s)|k>=\frac{v^2(k)}{{\cal I}(s)}
\end{equation}
In the Unitary limit ${\cal I}(s)$ is given by eq. (\ref{UDG1})
with  $s<0$ in the Faddeev equation, eq. (\ref{unifadd}).
But for   free space
2-body scattering one has $s=k^2$ so that with momenta $k<1$
(in units of $\Lambda$) one will have $0<s<1$.
The integral, can again be done analytically and yields $${\cal I}(0<s<1)=0$$
so that the Reactance matrix
element $$<k|{\cal R}|k>=\frac{1}{k}\tan \delta(k)\rightarrow \infty$$ i.e.
$\delta(k)=\frac{\pi}{2}$ for ALL momenta $0<k<1$ and the condition for a
unitary interaction with a cut-off $\Lambda$ in momentum-space is
satisfied.

\end{document}